\documentstyle[12pt,aaspp4]{article}

\newcommand{\uc}{\rm U Com~}

\newcommand{\lsun}{log $L/L_{\odot}\,$}
\newcommand{\msun}{$M/M_{\odot}\,$}

\begin{document}

\title{The RR Lyrae {\rm U Com} as a test for nonlinear pulsation models}

\author{Giuseppe Bono\altaffilmark{1}, Vittorio Castellani\altaffilmark{2}, 
Marcella Marconi\altaffilmark{3}}

\affil{1. Osservatorio Astronomico di Roma, Via Frascati 33,
00040 Monte Porzio Catone, Italy; bono@coma.mporzio.astro.it}
\affil{2. Dipartimento di Fisica Universit\`a di Pisa, Piazza Torricelli 2,
56100 Pisa, Italy; vittorio@astr18pi.difi.unipi.it}
\affil{3. Osservatorio Astronomico di Capodimonte, Via Moiariello 16,
80131 Napoli, Italy; marcella@na.astro.it}

\begin{abstract}
We use high precision multiband photometric data of the first
overtone RR Lyrae \uc to investigate the predictive capability 
of full amplitude, nonlinear, convective hydrodynamical models.  
The main outcome of this investigation is that theoretical predictions 
properly account for the luminosity variations along a full pulsation 
cycle. Moreover, we find that this approach, due to the strong dependence 
of this observable and of the pulsation period on stellar parameters, 
supply tight constraints on stellar mass, effective temperature, and 
distance modulus. Pulsational estimates of these parameters appear 
in good agreement with empirical ones.
Finally, the occurrence of a well-defined bump just before the luminosity 
maximum gave the unique opportunity to calibrate the turbulent convection 
model adopted for handling the coupling between pulsation and convection.  
\end{abstract}


%%%%%%%%%%%%%%%%%%%%%%%%%%%%%%%%%%%%%%%%%%%%%%%%%%%%%%%%%%%%%%%%%%%%%%%%%%%%%%%
\pagebreak 
\section{Introduction} 
 
Variable stars play a key role in many astrophysical problems, 
since their pulsation properties do depend on stellar parameters, 
and therefore they can supply valuable and independent constraints 
on a large amount of current evolutionary predictions. 
In particular, the empirical evidence found long time ago in 
Magellanic Cepheids of the correlation between period and 
luminosity was the initial step for a paramount theoretical and 
observational effort aimed at using variable stars as standard 
candles to estimate cosmic distances. The current literature is still 
hosting a vivid debate on the intrinsic accuracy of the Cepheid distance 
scale (Bono et al. 1999; Laney 2000) and on the use of RR Lyrae 
variables to evaluate the distance -and the age- of Galactic globulars  
(Caputo 1998; Gratton 1998, G98). 

Theoretical insights into the problem of radial stellar pulsations 
came from the linearization of local conservation equations governing 
the dynamical instability of stellar envelopes. Linear, nonadiabatic 
models typically supply accurate pulsation periods and plausible 
estimates (necessary conditions) on the modal stability of the 
lowest radial modes. 
However, a proper treatment of radial pulsations does require the 
solution of the full system of hydrodynamic equations, including 
a nonlocal and time-dependent treatment of turbulent convection 
(TC) to account for the coupling between radial and convective 
motions (Castor 1968; Stellingwerf 1982, S82). 
 
The development of nonlinear, convective hydrocodes (S82; Gehmeyr 1992; 
Bono \& Stellingwerf 1994, BS; Wuchterl \& Feuchtinger 1998) 
gave the opportunity to provide plausible predictions on the properties 
of radial variables, and in particular on the topology of the instability 
strip, as well as on the time behavior of both light and radial velocity 
curves. This new theoretical scenario allowed to investigate, for the 
first time, the dependence of pulsation amplitudes and Fourier 
parameters on stellar mass, luminosity and effective temperature 
(see e.g. Kovacs \& Kanbur 1997; Brocato et al. 1996; Feuchtinger 2000, F20).
However, all these investigations dealt with parameters related to 
the light curve, whereas nonlinear computations supply much more 
information, as given by the detailed predictions of the light 
variation along a full pulsation cycle. Therefore, the direct 
comparison between observed and predicted light curves appears as 
a key test only partially exploited in the current literature 
(Wood, Arnold, \& Sebo 1997, WAS). 

In order to perform a detailed test of the predictive capability of our 
nonlinear, convective models we focused our attention on the photometric  
data collected by Heiser (1996, H96) for the field, first 
overtone -$RR_c$- variable \uc. The reason for this choice relies 
on the detailed coverage of the U, B, and V light curves, as well as   
on the characteristic shape of the light curve, with a well-defined 
bump close to the luminosity maximum. This secondary feature provides 
a tight observational constraint to be nailed down by theory.  
Since the period of the variable strongly depends on the structural 
parameters (mass, luminosity, and radius) of the pulsator, 
the problem arises whether or not nonlinear pulsation models 
account for the occurrence of similar pulsators, and in affirmative 
how precisely the observed light curves can be reproduced by 
theoretical predictions.

In \S 2 we present the comparison between theory and observations, 
while in \S 3 we discuss the calibration of the TC model. Finally,
in \S 4 we briefly outline the observables which can further 
validate this theoretical scenario.

%%%%%%%%%%%%%%%%%%%%%%%%%%%%%%%%%%%%%%%%%%%%%%%%%%%%%%%%%%%%%%%%%%%%%%%%%%%%

\section{Comparison between theory and observations}
%%%%%%%%%%%%%%%%%%%%%%%%%%%%%%%%%%%%%%%%%%%%%%%%%%%%%%%%%%%%%%%%%%%%%%%%%%%%

On the basis of spectroscopic measurements Fernley \& Barnes (1997, FB97) 
estimated for \uc a metallicity $[Fe/H]=-1.25\pm0.20$, 
while Fernley et al. (1998, FB98) found a negligible interstellar 
extinction ($E(B-V)=0.015\pm0.015$).  
According to these empirical evidence and to a well-established 
evolutionary scenario, we expect for a metal-poor RR Lyrae a 
stellar mass of the order of 0.6 $M_\odot$ and a luminosity 
ranging from \lsun=1.6 to 1.7. At the same time, pulsation 
predictions on double-mode pulsators suggest similar mass values 
(Cox 1991; Bono et al. 1996). 
As a consequence, we computed a sequence of nonlinear models at fixed 
chemical composition (Y=0.24, Z=0.001) and pulsation period 
$P_{FO}\approx0.29$ d. 
Along such an iso-period sequence the individual models were constructed 
at fixed mass value (\msun = 0.60), while both the luminosity and the 
effective temperature were changed according to the pulsation relation 
given by Bono et al. (1997, BCCM). 
Both linear and nonlinear models were computed by adopting the input
physics, and physical assumptions already discussed in BS, BCCM, 
and in Bono, Marconi \& Stellingwerf (1999). According to S82, 
current models were computed by assuming a vanishing efficiency 
of turbulent overshooting in the region where the superadiabatic 
gradient attains negative values. This means that the convective flux 
can only attain positive or vanishing values ($F_c \ge 0$). 
In the next section we show that the assumption adopted 
in our previous investigations -i.e. $F_c$ can attain both 
positive and negative values- marginally affects the topology 
of the instability strip, but the predicted light curves are 
somewhat at variance with empirical ones.    

The top panels of Fig. 1 show that at fixed stellar mass (\msun=0.60), 
the double peak feature appears in models characterized by 
luminosities approximately equal to \lsun$\approx$1.61, and effective 
temperatures ranging from 6950 to 7150 K. These models also present 
{\rm B} amplitudes in reasonable agreement with empirical estimates 
($A_B=0.64$ mag). The bottom panels of Fig. 1 display the light 
curves of models 
along the iso-period sequence constructed by adopting a fixed effective 
temperature ($T_e=7100$ K) but different assumptions on stellar mass 
and luminosity. A glance at these curves shows that the luminosity 
amplitude is mainly governed by the stellar mass, whereas the shape 
of the light curve is only marginally dependent on this parameter. 

We find that the best fit to the observed B light curve is obtained 
for \msun=0.6, \lsun=1.607, $T_e=7100$ K, $P_{FO}=0.290$ d, together 
with a distance modulus $(m_B-M_B) = 11.01$ mag. The fit -though not perfect- 
appears rather satisfactory, thus suppling a substantial support 
to the predictive impact of the adopted theoretical scenario.
On the basis of this finding, we are now interested in testing the 
accuracy of theoretical predictions in different photometric bands. 
Fig. 2 shows from left to right the comparison between predicted 
light curves (solid lines) and empirical data (open circles) in the 
U, B, V, and K band respectively. The comparison was performed by 
adopting the same distance modulus, i.e. by neglecting 
the interstellar extinction, and the agreement between theory and 
observations seems even better than for the B light curve.  
Thus suggesting that nonlinear models account for luminosity 
amplitudes which are a long-standing problem of pulsation theory.

Not surprisingly, we also find that the time average colors predicted 
by our model appear, within current uncertainty on both  reddening and 
 photometry, in very good agreement with empirical estimates 
(see Table 1). This result supports the evidence that nonlinear models, 
at least in this case, can constrain stellar colors by best fitting 
the light curve in a single photometric band. At the same time,
this agreement suggests that the pulsational constraints on the 
temperature of the pulsator, as derived by the B light curve, are
consistent with the theoretical light curves in the other photometric 
bands.

\small 
\begin{center}
\begin{tabular}{cccc}
\tablewidth{0pt}\\  
\multicolumn{4}{c}{TABLE 1. {\rm U Com}: theoretical and empirical
colors$^a$}\\
\hline 
Color    & Theory        & Observ.      & Observ.          \\   
% mag      &              & $E_{(B-V)}=0$ & $E_{(B-V)}=0.015$ \\   
 mag      &              & $E(B-V)=0$ & $E(B-V)=0.015$ \\   
\hline 
$<U>-<B>$ &$0.06\pm0.01$ & $0.11\pm0.02$    & $0.09\pm0.02$  \\ 
$<B>-<V>$ &$0.23\pm0.01$ & $0.21\pm0.02$    & $0.19\pm0.02$  \\ 
$<V>-<K>$ &$0.77\pm0.01$ & $0.81\pm0.07$    & $0.77\pm0.07$  \\ 
\hline 
\end{tabular}
\end{center}
\normalsize 
\begin{minipage}{1.00\linewidth}  
\noindent $^a$ Empirical estimates are based on photometric 
data collected by H96 and by FB97. Theoretical colors refer to the best 
fit model, and the errors were estimated by assuming an uncertainty of 
50 K in the temperature of this model.  
\end{minipage} 

However, we note that on the basis of both the period and the 
shape of the {\rm B} light curve we predicted the effective temperature, 
the intrinsic luminosity, and in turn the distance modulus of this object. 
The plausibility of the theoretical constraints can be further tested
by comparing them with independent evaluations available in the literature. 
We find that the effective temperature predicted by nonlinear models 
-$T_e=7100\pm50$ K- is in remarkable agreement with the empirical 
temperature -$T_e=7100\pm150$ K- derived by adopting the true intensity 
mean color $(<V>-<K>)_0=0.77\pm0.07$ provided by FB97 and the 
CT relation by Fernley (1989). The same outcome applies by assuming 
E(B-V)=0, and indeed 
$(<V>-<K>)=0.81\pm0.07\,\rightarrow\,T_e=7050\pm150$ K, while the 
semi-empirical estimate provided by H96 suggests $T_e=7250\pm150$ K. 
We also note that the effective gravity of the best fit model 
(log $g\approx3.0$) is also in very good agreement with both the 
photometric estimate obtained by H96 (log $g=3.1\pm0.2$) and the 
spectroscopic measurements for field RR Lyrae variables provided 
by Clementini et al. (1995) and by Lambert et al. (1996). 

As far as the distance modulus is concerned, Fig. 3 shows the comparison 
of our pulsational estimates (filled circles) with empirical and 
theoretical \uc absolute magnitudes obtained by adopting different 
methods. The top and the bottom panel refer to absolute magnitudes  
based on visual and NIR magnitudes respectively. Data plotted in 
this figure show that our estimates appear in satisfactory  
agreement with distances based on the Baade-Wesselink method,  
on the statistical parallax method, on Hipparcos trigonometric 
parallaxes and proper motions, and on RR Lyrae {\rm K} band 
PL relation. 

However, one also finds that the distance determination based on 
HB evolutionary models constructed by including the most recent
input physics (Cassisi et al. 1999, C99) seems to overestimate 
the distance 
modulus by approximately 0.18 mag when compared with the current 
pulsation determination. A disagreement between evolutionary and 
pulsation predictions concerning the luminosities of RR Lyrae 
stars was brought out by Caputo et al. (1999), and more recently by 
Castellani et al. (2000) who found that up-to-date
He-burning models seem too bright when compared with Hipparcos 
absolute magnitudes. Data plotted in Fig. 3 confirm this discrepancy
between evolutionary and pulsational predictions, possibly due to 
an overluminosity of HB models. Part of this discrepancy might be 
due to the higher temperature of U Com when compared with the mean 
temperature of RR Lyrae gap ($T_e=6800$ K) adopted in evolutionary 
estimates.  

Finally, we also constructed several sequences of models by increasing
or decreasing the metal abundance by a factor of two. We find that the 
bump becomes more (less) evident at lower (higher) metal contents, and 
that this change does not allow us to obtain a good fit between 
theory and observations. This result is in satisfactory  agreement with 
the spectroscopic estimate by FB97 and supports the evidence that the 
shape of the light curve can also be used to constrain the $RR_c$ 
metallicity (Bono, Incerpi, \& Marconi 1996).

%%%%%%%%%%%%%%%%%%%%%%%%%%%%%%%%%%%%%%%%%%%%%%%%%%%%%%%%%%%%%%%%%%%%%%%%%%%%
\section{Calibration of the TC model}

The first set of models we constructed for fitting the empirical 
light curves was characterized by an unpleasant feature: 
the peak of the bump was, in contrast with empirical evidence, brighter
than the "true" luminosity maximum. According to Bono \& Stellingwerf
(1993) the bump along the rising branch presents a strong dependence 
on the free parameters adopted in the TC model. However, in the 
calibration of the TC model suggested by BS both the eddy viscosity 
and the diffusion scale lengths (see their equ. 4 and 9) were scaled 
to the value of the mixing length parameter. We performed several 
numerical experiments by 
changing along each sequence only one of the three free parameters. 
As a result, we find that full amplitudes models constructed by adopting 
plausible changes of the free parameters do not simultaneously account 
for the pulsation amplitude, the shape of the light curve, and the 
temperature width of first overtone instability region. 

Due to the lack of a self-consistent theory of time-dependent, nonlocal, 
convective transport, current investigations were mainly aimed at 
calibrating the free parameters adopted for treating the coupling 
between pulsation and convection (Yecko et al. 1998; F20). This is not 
a trivial effort, since the observables and the comparison between theory 
and observations are affected by the thorny problem of the transformation 
into the observational plane and/or by systematic deceptive errors 
such as reddening and distance estimates. 
In order to overcome some of these difficulties, F20 calibrated  
the TC model by performing a detailed comparison between theoretical 
and observed luminosity amplitudes of field RR Lyrae variables. On the basis 
of the fine tuning of both mixing length and turbulent viscosity length, 
F20 found that the Fourier parameters of fundamental light curves agree 
with observational data. The same outcome did not apply to $RR_c$ 
variables, and indeed predicted values appear, at fixed period, 
smaller than the empirical ones.  
A detailed comparison with the convective 
structure of RR Lyrae models constructed by F20 is not possible because 
he adopted a convective flux limiter in the turbulent source function 
and in the convective flux enthalpy, and neglected both the turbulent 
pressure and the turbulent overshooting. As a consequence, we decided 
to test the dependence of full amplitude models on the last two 
ingredients. 

Interestingly enough, we find that the models constructed
by assuming a vanishing overshooting efficiency satisfy empirical 
constraints i.e. the bump is dimmer than the luminosity maximum, the 
luminosity amplitudes attain values similar to the observed ones 
and the temperature width of the region in which the first overtone 
is unstable agrees with BCCM findings for cluster $RR_c$ variables. 
Fig. 4 shows the 
{\rm B} light curve of two models constructed by adopting 
the same input parameters, but different assumptions on the efficiency 
of turbulent overshooting. It is noteworthy the simultaneous change in 
the pulsation amplitude and in the shape of the light curve. 
The dependence of the convective structure on both overshooting 
and turbulent pressure will be investigated in a forthcoming paper.

%%%%%%%%%%%%%%%%%%%%%%%%%%%%%%%%%%%%%%%%%%%%%%%%%%%%%%%%%%%%%%%%%%%%%%%%%%%%
\section{Discussion and Conclusions} 

The comparison between theory and observations, namely the period and 
the shape of the {\rm B} light curve of U Com, allowed us to supply 
tight constraints on the structural parameters such as stellar mass, 
effective temperature, and gravity, as well as on the distance of this 
variable. We also found that the occurrence of a well-defined bump 
close to the luminosity maximum can be safely adopted for 
constraining the metallicity of this object and for calibrating 
the TC model adopted for handling the coupling between convection 
and pulsation. 

The approach adopted in this investigation seems quite promising since 
it only relies on nonlinear, convective models and on stellar atmosphere 
models. In fact, the best fit model to the empirical data was found by 
constructing sequences of iso-period models in which the stellar mass, 
the luminosity and the effective temperatures were not changed according 
to HB models but to the pulsation relation (BCCM).  
The comparison between theory and observations  
shows that both the structural parameters and the distance are 
in very good agreement with estimates available in the literature.  
No evidence for a systematic discrepancy was found in pulsation estimates, 
thus supporting the evidence that the individual fit to light curves  
can supply independent and firm constraints on the actual parameters 
and distances of variable stars. This finding confirms the results 
of a similar analysis on a LMC Bump Cepheid by WAS.  

Accurate radial velocity data for \uc are not available in the 
literature and therefore we could not constrain the accuracy of the 
velocity variation along the pulsation cycle. The three radial velocity 
points collected by FB98 agree quite well with the 
predicted curve. However, the radial velocity curve is a key 
observable for constraining the consistency of the adopted TC 
model (F20), and therefore new spectroscopic measurements 
of \uc would be of great relevance for assessing the 
predictive impact of nonlinear, convective models. 
Theoretical observables of the best fit model discussed in this 
paper, as well as both radius and radial velocity variations 
are available upon request to the authors.

%\acknowledgments
It is a pleasure to thank M. Groenewegen for providing us the
\uc distances based on the reduced parallax method and on 
the modified Lutz-Kelker correction, as well as for insightful 
discussions on their accuracy. We are indebted to R. Garrido 
for sending us radial velocity data and to T. Barnes for useful 
suggestions on current data. We also acknowledge an anonymous 
referee for some useful suggestions that improved the readability 
of the paper. This work was supported by MURST -Cofin98- under the 
project "Stellar Evolution". Partial support by ASI and CNAA is 
also acknowledged.

%%%%%%%%%%%%%%%%%%%%%%%%%%%%%%%%%%%%%%%%%%%%%%%%%%%%%%%%%%%%%%%%%%%%%%%%
%                           Figure Captions
%%%%%%%%%%%%%%%%%%%%%%%%%%%%%%%%%%%%%%%%%%%%%%%%%%%%%%%%%%%%%%%%%%%%%%%%
\pagebreak
% 1
\figcaption{Top panels: blue light curves of iso-period -$P=0.29$ d- RR Lyrae 
models constructed by adopting the same chemical composition (Y=0.24, 
Z=0.001) and stellar mass (\msun=0.60), but different assumptions on 
effective temperatures and luminosities (see labeled values). 
Bolometric curves where transformed into {\rm B} magnitudes by 
adopting bolometric corrections and color-temperature (CT) relations 
by Castelli, Gratton \& Kurucz (1997).  
Bottom panels: similar to top panels, but for models constructed at 
fixed effective temperature ($T_e=7100$ K), and different assumptions 
on stellar masses and luminosities (see labeled values).}

% 2
\figcaption{Comparison between theory (solid lines) and observations 
(open circles). From left to right the panels refer to photometric 
data in U, B, V (Heiser 1996), and K (Fernley, Skillen \& Burki 1993).  
Empirical data were plotted by assuming $E(B-V)=0$. Photometric errors 
in optical bands are equal to the symbol size.}  

% 3 
\figcaption{\uc absolute magnitudes vs. time. Top and bottom panels 
show {\rm V} and {\rm K} estimates respectively. The acronyms refer 
to different methods: 
BW: Baade-Wesselink (Jones et al. 1992); HB models: Horizontal Branch
models (Cassisi et al. 1999, C99); SP: statistical parallax
(Barnes \& Hawley 1986, BH); TP: trigonometric parallax (Gratton 1998, G98);
BW + PLK: zero point from BW and slope from {\rm K} band Period-Luminosity
relation (Longmore et al. 1990, L90; J92; Carney et al. 1995, C95);
LK + BW: zero point from modified Lutz-Kelker correction (Groenewegen \&
Salaris 1999, GS) and slope from BW (Fernley et al. 1998, FC98);
RP + BW: zero point from reduced parallax (GS) and slope from BW (FC98);
SP + BW: zero point from SP (Layden et al. 1996, L96; Fernley et al. 1998,
FB98; Gould \& Popowski 1998, GP; Tsujimoto, Miyamoto, \& Yoshii 1998, TMY;
Layden 1999, L99) and slope from BW (FC98).} 

% 4 
\figcaption{ {\rm B} light curves of RR Lyrae models vs. phase. 
Solid lines refer to models constructed by adopting the calibration 
of the TC model suggested by BS, while dashed lines to models 
constructed by assuming that the convective flux is vanishing in the 
regions in which the superadiabatic gradient is negative 
(see equ. 7 and \S 3 in BS).}  
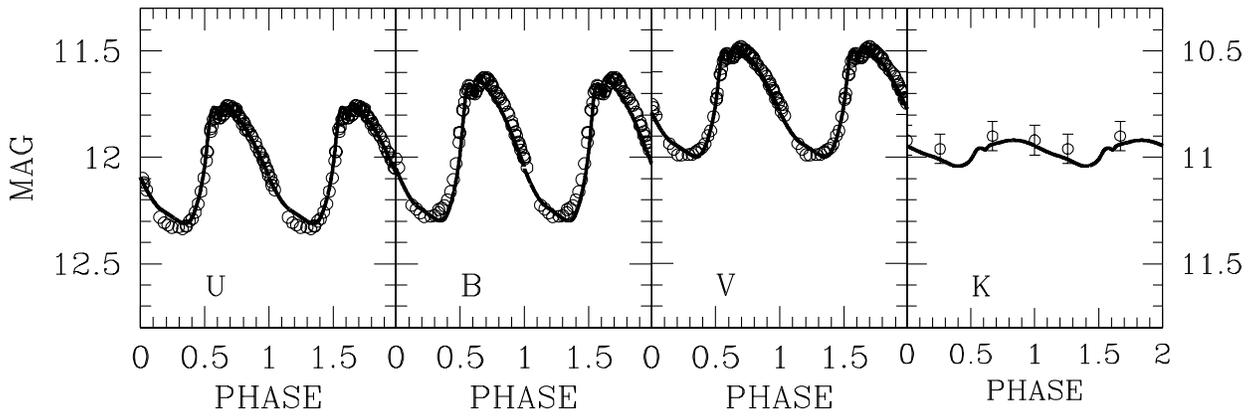
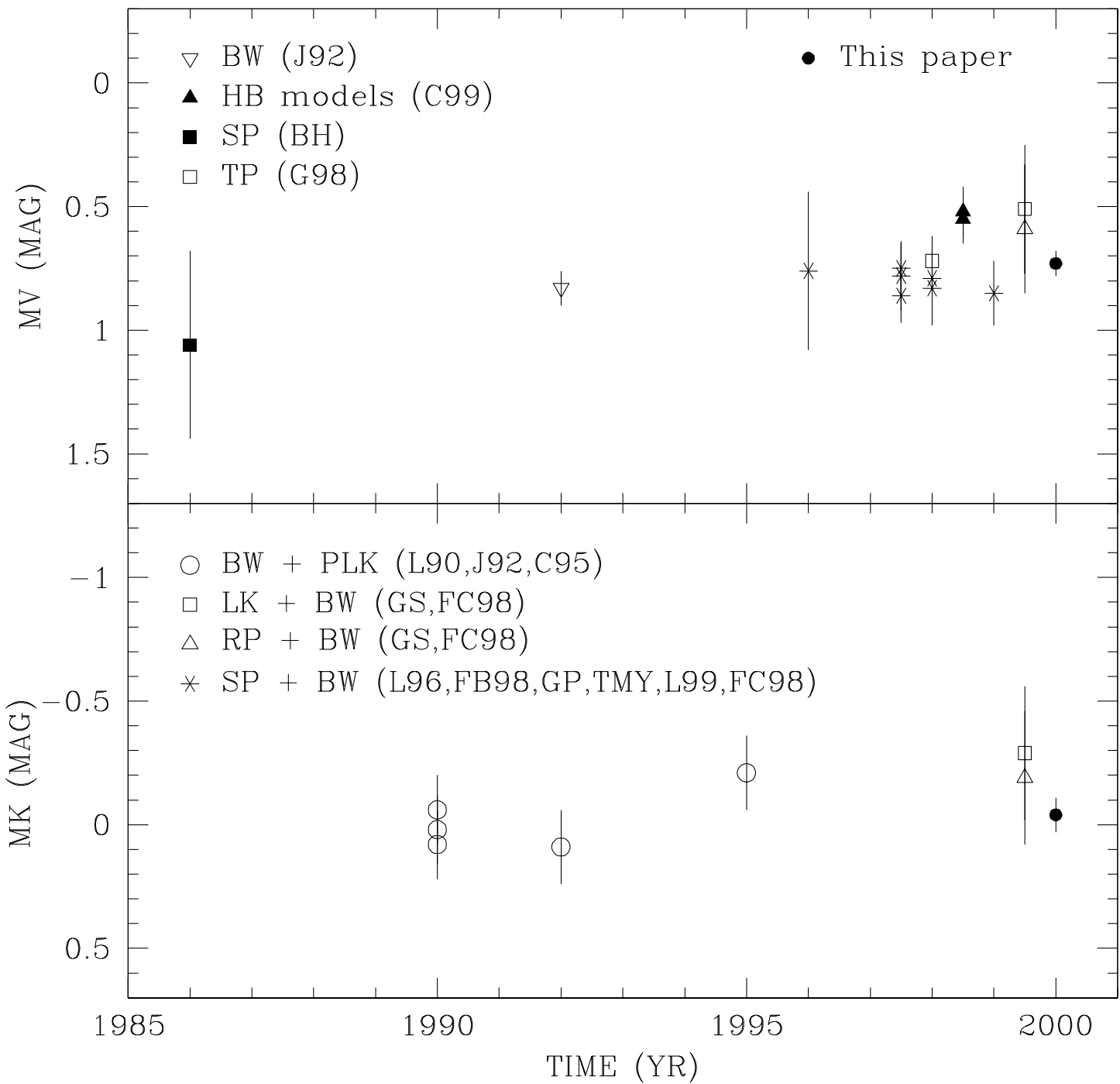
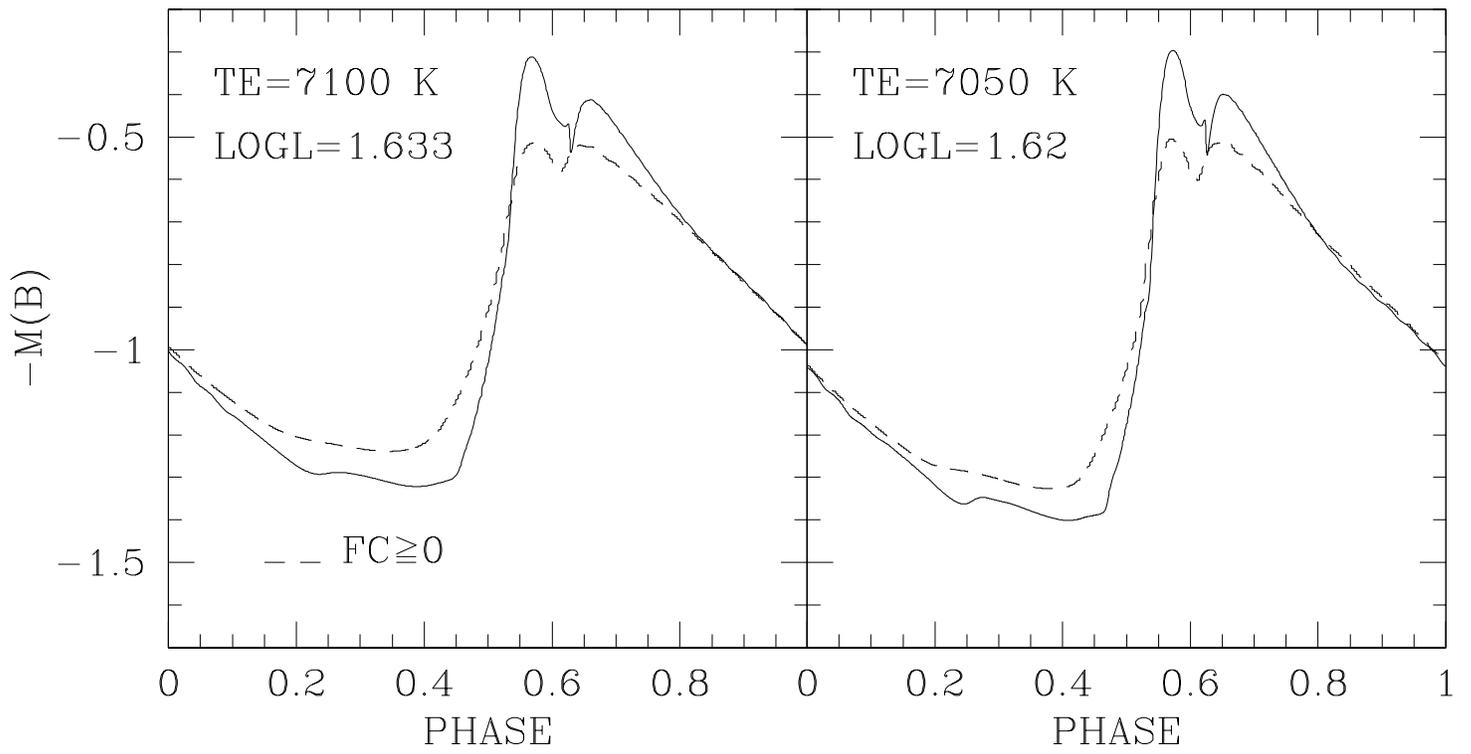
\end{document}